\documentclass[prl,amsmath,aps,twocolumn,superscriptaddress,showpacs]{revtex4}
\usepackage{amsfonts}
\usepackage{graphicx}
\usepackage{times}

\begin{document}
\title{Joint measurement of two unsharp observables of a qubit}
\author{Sixia Yu}
\affiliation{Hefei National Laboratory for Physical Sciences at
Microscale and Department of Modern Physics, University of Science
and Technology of China, Hefei, Anhui 230026, P.R. China}
\affiliation{Centre for Quantum Technologies and Physics Department,
National University of Singapore, 2 Science Drive 3, Singapore
117542}
\author{Nai-le Liu}
\affiliation{Hefei National Laboratory for Physical Sciences at
Microscale and Department of Modern Physics, University of Science
and Technology of China, Hefei, Anhui 230026, P.R. China}
\author{Li Li}
\affiliation{Hefei National Laboratory for Physical Sciences at
Microscale and Department of Modern Physics, University of Science
and Technology of China, Hefei, Anhui 230026, P.R. China}
\author{C. H. Oh}
\affiliation{Centre for Quantum Technologies and Physics Department,
National University of Singapore, 2 Science Drive 3, Singapore
117542}

\begin{abstract}
We present a single inequality as the necessary and sufficient
condition for two unsharp observables of a two-level system to be
jointly measurable in a single apparatus and construct explicitly
the joint observables. A complementarity inequality arising from the
condition of joint measurement, which generalizes Englert's duality
inequality,  is derived as the trade-off between the unsharpnesses
of two jointly measurable observables.
\end{abstract}
\pacs{03.65.Ta, 03.67.-a}

\maketitle

Built in the standard formalism of quantum mechanics, there are
mutually exclusive but equally real  aspects of quantum systems, as
summarized by the complementarity principle of Bohr \cite{bohr}.
Mutually exclusive aspects are often exhibited via noncommuting
observables, for which the complementarity is quantitatively
characterized by two kinds of uncertainty relationships, namely, the
preparation uncertainty relationships (PURs) and the measurement
uncertainty relationships (MURs).

The PURs stem from the semi positive definiteness of the density
matrix describing the quantum state and characterize the
predictability of two noncommuting observables in a given quantum
state. To test PURs two different projective measurements will be
performed on two identically prepared ensembles of the quantum
system and these measurements cannot be performed within one
experimental setup on a single ensemble.

On the other hand MURs characterize the trade-off between the
precisions of unsharp measurements of two noncommuting observables
in a single experimental setup. The very first effort of Heisenberg
\cite{heis} in deriving the uncertainty relationships was based on a
simultaneous measurement of the position and momentum, with the
rigorous form of MUR  established recently by Werner \cite{werner}.
In the interferometry the wave-particle duality between the
path-information and the fringe visibility of interference pattern
is characterized quantitatively by Englert's duality inequality
\cite{berg}, which turns out to be originated from the joint
measurability of two special unsharp observables encoding the path
information and the fringe visibility \cite{liu}. To establish a
general MUR the condition for joint measurement  has to be explored, which can be turned into some kinds
of MURs when equipped with proper measure of the precisions (e.g.,
distinguishability).

In this Letter we shall consider the joint measurability of two
general unsharp observables of a qubit and derive a simple necessary
and sufficient condition with joint observables explicitly
constructed. We also present a MUR arising from the condition of
joint measurement that generalizes Englert's duality inequality.

{\em Joint measurability --- }Generally an observable is described
by a positive-operator valued measure (POVM), a set of positive
operators $\{O_k\}_{k=1}^K$ summed up to the identity ($O_k\ge0$ and
$\sum_kO_k=I$) with $K$ being the number of outcomes. By definition,
a {\it joint measurement} of two observables $\{O_k\}$ and
$\{O_l^\prime\}$ is described by a {\it joint observable}
$\{M_{kl}\}$ whose outcomes can be so grouped that
\begin{equation}\label{me}
O_k=\sum_l M_{kl},\quad O_{l}^\prime=\sum_k M_{kl}.
\end{equation}

Here we shall consider the qubits, any two-level systems such as
spin-half systems or two-path interferometries. A {\it simple
observable} $\mathcal O(x,\vec m)$ refers to a most general
2-outcome POVM $\{O_\pm(x,\vec m)\}$ with
\begin{equation}
O_\pm(x,\vec m)=\frac{1\pm (x+\vec m\cdot \vec \sigma)}2.\label{o}
\end{equation}
Here $m=|\vec m|$ is referred to as the \textit{sharpness} while
$|x|$ is referred to as the \textit{ biasedness}. When $|x|=0$ the
observable $\mathcal O(x,\vec m)$ is called as {\it unbiased}, in
which case the outcomes of measurement are purely random if the
system is in the maximally mixed state, and when $|x|\neq0$ the
observable is referred to as {\it biased}, in which case priori
information can be employed to make better use of the outcomes of
the measurement. Positivity imposes  $|x|+m\le 1$.

Given two simple observables  $\mathcal O(x,\vec m)$ and $\mathcal
O(y,\vec n)$, it is obvious that all possible sets of four operators
satisfying the marginal constraints Eq.~(\ref{me}) are
\begin{equation}\label{m1}
M_{\mu\nu}(Z,\vec z)=\frac{1+\mu x+\nu y+\mu\nu Z+(\mu\nu\vec z+\vec
q_{\mu\nu})\cdot\vec \sigma}{4}
\end{equation}
with $Z,\vec z$ being arbitrary and $\vec q_{\mu\nu}=\mu\vec
m+\nu\vec n$ $(\mu,\nu=\pm1)$. The problem of joint measurability
becomes whether there exist $Z,\vec z$ such that $M_{\mu\nu}(Z,\vec
z)\ge0$ for all $\mu,\nu=\pm1$. There are many partial results in
special cases \cite{b1,liu,b3} as well as in general cases
\cite{b2,srh}. Here we shall present a single inequality as the
condition. For convenience we denote
\begin{subequations}
\begin{eqnarray}
F_x&=&\frac12\left(\sqrt{(1+ x)^2-m^2}+\sqrt{(1- x)^2-m^2}\right),\\
F_y&=&\frac12\left(\sqrt{(1+ y)^2-n^2}+\sqrt{(1- y)^2-n^2}\right).
\end{eqnarray}
\end{subequations}

{\bf Theorem 1} Two observables $\mathcal O(x,\vec m)$ and $\mathcal
O(y,\vec n)$ are jointly measurable iff
\begin{equation}\label{one}
(1-F_x^2-F_y^2)\left(1-\frac{x^2}{F_x^2}-\frac{y^2}{F_y^2}\right)\le
(\vec m\cdot\vec n-xy)^2.
\end{equation}

Due to the identities such as $x^2/F_x^2+m^2/(1-F_x^2)=1$ the
left-hand-side of Eq.~(\ref{one}) can be seen to be bounded above by
$(mn-|xy|)^2$ so that the trivial case $s=0$ with $s=|\vec
m\times\vec n|$ is included. In the case of $x=y=0$ Eq.~(\ref{one})
reproduces the condition $m^2+n^2\le1+(\vec m\cdot\vec n)^2$  for
unbiased observables \cite{b1}. When $y=0$ the condition
Eq.~(\ref{one}) reads $F_x\sqrt{m^2-(\vec m\cdot\vec n)^2}\ge s$
which becomes simply $F_x\ge n$ for orthogonal observables where
$\vec m\cdot\vec n=0$ \cite{liu}.

More generally we refer a pair of observables that satisfy
$\gamma=0$ where $\gamma=\vec m\cdot\vec n-xy$ to as {\em orthogonal
unsharp observables}. The condition of joint measurement
Eq.~(\ref{one}) becomes simply $F_x^2+F_y^2\ge1$ because
${x^2}/{F_x^2}+{y^2}/{F_y^2}< 1$ is ensured by $mn>|xy|$. In general
the condition $F_x^2+F_y^2\ge1$ is sufficient for joint measurement
since Eq.~(\ref{one}) is ensured because $(|xy|-mn)^2\le \gamma^2$
when $mn<xy$. Specifically we refer a pair of observables that
satisfy $\gamma=0$ and $\vec n=\vec n_c$ with
$n^2_c/F_x^2+y^2/(1-F_x^2)=1$ to as a pair of {\em maximally
orthogonal unsharp observables}. It is maximal because any
observable $\mathcal O(y,\vec n)$ with $n\le n_c$ (regardless of its
direction) is jointly measurable with $\mathcal O(x,\vec m)$ while
all the observables $\mathcal O(y,\vec n)$ with $n>n_c$ along $\vec
n_c$ are not jointly measurable with $\mathcal O(x,\vec m)$.

As the measure for unsharpness  we take a linear combination of the
sharpness and the biasedness , i.e., $D_1=Q_1 m+P_1|x|$ and
$D_2=Q_2n+P_2|y|$ with $0\le P_i\le Q_i$ $(i=1,2)$ being some
constants. To measure jointly a pair of orthogonal unsharp
observables there is a trade-off between the above defined
unsharpnesses (since $D_1^2+(Q_1^2-P_1^2)F_x^2\le Q_1^2$)
\begin{equation}
{D_1^2}{(Q_2^2-P_2^2)}+{D_2^2}{(Q_1^2-P_1^2)}+P_1^2P_2^2\le
Q_1^2Q_2^2.
\end{equation}
Englert's duality inequality \cite{berg} in the case of orthogonal
observables with one being unbiased \cite{liu} turns out to be a
special case of the above inequality if we let $Q_1=1$, $P_{2}=0$ so
that $D_1$ and $D_2$ become the path distinguishability and the
fringe visibility respectively. Theorem 1 is derived from the
following set of conditions.

{\bf Theorem 2} Two observables $\mathcal O(x,\vec m)$ and $\mathcal
O(y,\vec n)$ $(\vec m\times\vec n\neq 0)$ are jointly measurable iff
either $\max\{|\alpha|,|\beta|\}\ge1$ or
\begin{eqnarray}\label{t}
\sum_{\nu=\pm}\left|\vec m+\vec n+\nu\vec
g\right|+\sum_{\nu=\pm}\left|\vec m-\vec n+\nu\vec g\right|\le4,
\end{eqnarray}
with $\vec g=\vec m\alpha+\vec n\beta$ and
\begin{subequations}
\begin{eqnarray}
\alpha=\frac1{|\vec m\times\vec n|^2}\left[(y+\gamma x)n^2-(x+\gamma y)\vec m\cdot\vec n\right],\\
\beta=\frac1{|\vec m\times\vec n|^2}\left[(x+\gamma y)m^2-(y+\gamma
x)\vec m\cdot\vec n\right].&
\end{eqnarray}
\end{subequations}

Now let us examine the set of all observables $\mathcal O(y,\vec n)$
with a given biasedness $y$ that are jointly measurable with a fixed
observable $\mathcal O(x,\vec m)$. The admissible region of  $\vec
n$ is shown in Fig.~1(a) as the union of a red- and a blue-contoured
regions with boundary given by Eq.~(\ref{one}) with equality and
$|y|+n=1$. The (blue) arcs of the circle $n=1-|y|$ satisfying
Eq.~(\ref{one}) define  a forward and a backward admissible cones
around $\vec m$ (centered on the origin) in which all $\vec n$ are
admissible. If ${1-F_x^2}\le |y|$ then Eq.~(\ref{one}) holds true
and two cones overlap so that all $\vec n$ are admissible as
formulated as one part of conditions in \cite{srh}. From Lemma~3~iv)
we see that Eq.~(\ref{t}) is equivalent to $R\ge0$ with
\begin{eqnarray} R=1+x^2+y^2+\gamma^2-m^2-n^2-|\vec
g|^2\label{t1}
\end{eqnarray}
which appears in an equivalent form (Eq.~(55)) in \cite{b2}.

\begin{figure}
\includegraphics[width=80mm]{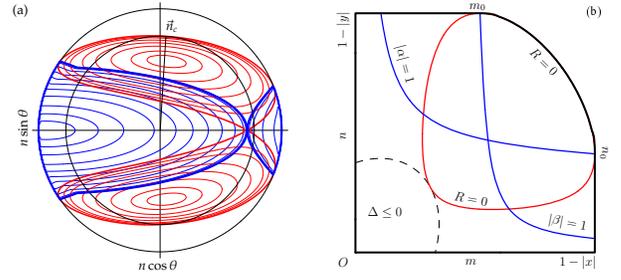}
\caption{(a) The union of the blue and red contoured regions,
determined by $\max\{|\alpha|,|\beta|\}\ge1$ and $R\ge0$
respectively, represents the admissible $\vec n$ in the case of
$m=0.8,x=-0.1$, and $y=0.3$. The boundary lies between two circles
$n=1-|y|$ and $n=n_c$. (b) The trade-off curve (solid black) between
sharpnesses $m$ and $n$ with $x=-0.1,y=0.2$, and $\cos\theta=0.3$
fixed.  }
\end{figure}

Taking the sharpnesses $m,n$ as measures for precisions we have
plotted their trade-off curve (solid black) in Fig.~1(b) with $x,y,$
and $\theta$ fixed where $\theta$ is the angle between $\vec m$ and
$\vec n$. There is a critical value $m_0$ of the sharpness
determined by Eq.~(\ref{one}) with equality and $|y|=1-n$, below
which there is no constraint on $n$. If
\begin{equation}
(1+\mbox{sgn}[xy]\cos\theta)(1-|x|)(1-|y|)\le 2|xy|,
\end{equation}
with ${\rm sgn}[f]=+1$ if $f\ge 0$ and $-1$ if $f<0$, then $m_0\ge
1-|x|$ (and $n_0\ge 1-|y|$) so that there is no trade-off between
$m,n$. Here $n_0$ is defined similarly as $m_0$ with $(x,m)$ and
$(y,n)$ interchanged. If $ m+n+|\vec m\pm\vec n|\le 2$ every vector
$\vec g=\vec m\alpha+\vec n\beta$ with
$\max\{|\alpha|,|\beta|\}\le1$ satisfies Eq.~(\ref{t}) so that there
is no trade-off between $x,y$.

{\em Joint unsharp observables --- }
If $s=0$ then a joint observable of observables $\mathcal O(x,\vec
m)$ and $\mathcal O(y,\vec n)$ is simply given by $\{O_\mu(x,\vec
m)O_{\nu}(y,\vec n)\}$. If $\Delta_\tau<0$ with $
\Delta_\tau=(\vec m-\tau\vec n)^2-(x-\tau y)^2 $
 for some $\tau=\pm$ then $O_{\eta}(x,\vec m)-O_{\eta\cdot\tau}(y,\vec
n)\ge0$ where $\eta={\rm sgn}[x-\tau y]$. Therefore the POVM
$
\{0,O_{\bar\eta}(x,\vec m),O_{\eta}(x,\vec
m)-O_{\eta\cdot\tau}(y,\vec n),O_{\eta\cdot\tau}(y,\vec n)\}
$
 is a joint observable ($\bar\eta=-\eta$). When $s>0$ and
$\Delta_\pm\ge 0$ we have:

{\bf Theorem 3}  Given observables $\mathcal O(x,\vec m)$ and
$\mathcal O(y,\vec n)$, a) if $R\ge0$ then $\{M_{\mu\nu}(\gamma,\vec
g)\}$ is a joint observable;  b) if $R<0$ and
$\max\{|\alpha|,|\beta|\}\ge1$ then $\{M_{\mu\nu}(Z(\vec
z_{\eta\tau}),\vec z_{\eta\tau})\}$ is a joint observable where
\begin{eqnarray}
&\displaystyle Z(\vec z)=\max_{\mu=\pm1}\left\{\left|\vec z+\mu(\vec
m+\vec n)\right|-\mu(x+y)\right\}-1,&\label{Z}\\
&\displaystyle\vec z_{\eta\tau}=\vec g+\frac{D_{\eta\tau}(\vec
m\times\vec n)\times\vec L_{\eta\tau}}{\vec L^2_{\eta\tau}-|\vec
m\times\vec n|^2},&
\end{eqnarray} with $D_{\eta\tau}=\tau A_\eta\alpha+\eta
B_\tau\beta+\eta\tau\gamma-1$, $\vec L_{\eta\tau}=\tau A_\eta\vec
n-\eta B_\tau\vec m$, $\tau={\rm sgn} [\alpha]$ and $ \eta={\rm
sgn}[B_\tau\beta +\tau\gamma-x]$ if $|\alpha|\ge 1$, and $\eta={\rm
sgn}[\beta]$ and $\tau={\rm sgn}[A_\eta\alpha+\eta\gamma-y]$ if
$|\beta|\ge1$, where $A_\eta=1-\eta x$ and $B_\tau=1-\tau y$.

The regions for different constructions of joint observables
according to the above theorem are indicated in Fig.~1(b) whenever
two observables are jointly measurable. We note that
$\Delta=\min\{\Delta_\pm\}<0$ infers $\max\{|\alpha|,|\beta|\}\ge1$.
Since $R=d^2_{\mu\nu}-|\vec g-q_{\mu\nu}|^2$ for all $\mu,\nu=\pm$
with $d_{\mu\nu}=1-\mu x-\nu
y+\mu\nu\gamma$, we see that if $R=0$ then $
4M_{\mu\nu}(\gamma,\vec g)={d_{\bar\mu\bar\nu}+(\mu\nu \vec g+\vec
q_{\mu\nu})\cdot\vec \sigma}$
 are proportional to some projections for all $\mu,\nu=\pm$.

{\em Proofs --- }We shall at first prove 3  relevant Lemmas in which
we make use of the fact that given two overlapping convex regions in
a plane either their boundaries intersect or one region belongs to
the other. In what follows we suppose  $s>0$. In the plane $\mathbb
P$ spanned by $\vec m$ and $\vec n$ we denote by
\begin{eqnarray}
\mathbb E^{\mu}_x&=&\textstyle\left\{\;\vec z\in \mathbb P\;\big|\;\sum_{\tau=\pm}|\vec z-\vec q_{\tau\mu}|\le 2(1-\mu x)\right\},\label{ex}\\
\mathbb E^{\nu}_y&=&\textstyle\left\{\;\vec z\in \mathbb
P\;\big|\;\sum_{\tau=\pm}|\vec z-\vec q_{\nu\tau}|\le 2(1-\nu
y)\right\}\label{ey}
\end{eqnarray}
four elliptical regions with boundaries being four ellipses
$E_x^\pm$ and $E_y^\pm$ whose semi-major and squared semi-minor axes
are denoted by $A_\mu=1-\mu x$, $B_\nu=1-\nu y$ and
$a_\mu=A_\mu^2-m^2$, $b_\nu=B_\nu^2-n^2$ respectively. Two
neighboring ellipses $E_x^\mu$ and $E_y^\nu$ have one focus
$Q_{\nu\mu}$ (corresponding to vector $\vec q_{\nu\mu}$) in common.
Also we denote by
\begin{equation}
\mathbb E=\left\{\;\vec z\in \mathbb P\;\big|\;\textstyle
\sum_{\mu,\nu=\pm}\left|\vec z-\vec q_{\mu\nu}\right|\le 4\right\}
\end{equation}
an oval region with boundary being a generalized ellipse $E$ with
four foci $Q_{\mu\nu}$ with $\mu,\nu=\pm$. The condition
Eq.~(\ref{t}) becomes $\vec g\in \mathbb E$. It is easy to see that
$\mathbb J_x:=\mathbb E_x^+\cap \mathbb E_x^-\subset \mathbb E$,
$\mathbb J_y:=\mathbb E_y^+\cap \mathbb E_y^-\subset \mathbb E$ with
boundaries given by $J_x=(E_x^+\cup E_x^-)\cap\mathbb E$ and
$J_y=(E_y^+\cup E_y^-)\cap\mathbb E$ respectively. Furthermore
$E_x^+\cap E_x^-\subset E\cap J_x$, $E_y^+\cap E_y^-\subset E\cap
J_y$.  In Fig.~2 two neighboring ellipses with intersections and the
4-ellipse $E$ (red curve) are shown.

\begin{figure}
{\includegraphics[width=45mm]{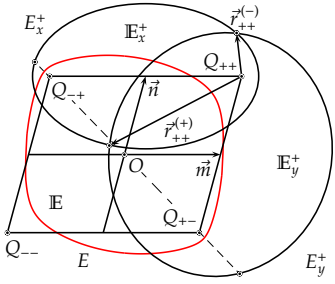}} \caption{ The setup for
proofs. In the plane $\mathbb P$ spanned by $\vec m$ and $\vec n$
there are two neighboring ellipses $E_x^+$ and $E_y^+$ and the
generalized ellipse $E$ with 4 foci $Q_{\mu\nu}$ (red curve).}
\end{figure}

{\bf Lemma 1}  Two observables $\mathcal O(x,\vec m)$ and $\mathcal
O(y,\vec n)$ are jointly measurable iff $ \mathbb J=\mathbb
E_x^+\cap\mathbb E_x^-\cap\mathbb E_y^+\cap \mathbb E_y^-\ne
\emptyset. $

{\bf Proof } If $\mathcal O(x,\vec m)$ and $\mathcal O(y,\vec n)$
are jointly measurable then there exist $Z$ and $\vec z$ such that
$M_{\mu\nu}(Z,\vec z)\ge0$, i.e.,
\begin{equation}\label{pc}
\left|\mu\nu\vec z+\vec q_{\mu\nu}\right|\le 1+\mu x+\nu
y+{\mu\nu}Z,
\end{equation}
for all $\mu,\nu=\pm$. As a result $\vec z-(\vec z\cdot \vec s)\vec
s/s^2\in\mathbb J$ with $\vec s=\vec m\times \vec n$. If there
exists $\vec z\in\mathbb J$ then Eq.~(\ref{pc}) holds true with $Z$
given by $Z(\vec z)$ as in Eq.~(\ref{Z}), i.e., $\{M_{\mu\nu}(Z(\vec
z),\vec z)\}$ is a joint observable.

{\bf Lemma 2} $\mathbb J\ne\emptyset$ iff either $ E_x^\mu\cap
E_y^\nu\cap\mathbb E\ne\emptyset$  for some $\mu,\nu=\pm$ or
$E_x^\mu\cap E_y^\nu=\emptyset$ for all $\mu,\nu=\pm$.

{\bf Proof } Sufficiency. Suppose that there exists $\vec z\in
E_x^\mu\cap E_y^\nu\cap \mathbb E$ for some $\mu,\nu=\pm$. From
$\vec z\in E_x^\mu \cap \mathbb E$ and $\vec z\in E_y^\nu\cap
\mathbb E$ it follows that $\vec z\in \mathbb E_x^{\bar\mu}$ and
$\vec z\in \mathbb E_y^{\bar\nu}$ respectively, which leads to $\vec
z\in \mathbb J$. If $E_x^\mu\cap E_y^\nu=\emptyset$ for all
$\mu,\nu$ then, taking into account of of Lemma 3 ii), we have
either $\mathbb E_x^\mu\subset\mathbb  E_y^\pm\subset \mathbb
E_x^{\bar\mu}$ or $\mathbb E_y^\nu\subset \mathbb E_x^\pm\subset
\mathbb E_y^{\bar\nu}$, i.e., either $\mathbb J=\mathbb E_x^\mu$ or
$\mathbb J=\mathbb E_y^\nu$ for some $\mu,\nu$, which is obviously
not empty.

Necessity. If $\mathbb J\ne \emptyset$ then two convex regions
$\mathbb J_x$ and $\mathbb J_y$ overlap. As a result we have either
$J_x\cap J_y\ne\emptyset$, which means $(\exists \mu,\nu)\
E_x^\mu\cap E_y^\nu\cap \mathbb E\ne\emptyset$, or $J_x\cap
J_y=\emptyset$ with either $\mathbb J_x\subset \mathbb J_y$ or
$\mathbb J_y\subset \mathbb J_x$. If $J_x\cap J_y=\emptyset$ and
$\mathbb J_x\subset \mathbb J_y$, i.e., $\mathbb J_x$ lies totally
within $\mathbb J_y\subset\mathbb E$, then $J_x\cap E=\emptyset$
which infers $E_x^+\cap E_x^-=\emptyset$, i.e., the boundaries of
two overlapping regions $\mathbb E_x^\pm$ do not intersect. As a
result $(\exists\mu)\ \mathbb E_x^\mu\subset \mathbb E_x^{\bar\mu}$
so that $(\exists\mu)\ \mathbb E_x^\mu=\mathbb J_x\subset\mathbb
E_y^\pm$, which infers $(\exists\mu)\ E_x^\mu\cap E_y^\pm=\emptyset$
since $J_x\cap J_y=\emptyset$. In the same manner $\mathbb
J_y\subset \mathbb J_x$ with $J_x\cap J_y=\emptyset$ infers
$(\exists\nu)\ E_y^\nu\cap E_x^\pm=\emptyset$. In both cases,
considering Lemma 3 i), we obtain $(\forall \mu,\nu)\ E_x^\mu\cap
E_y^\nu=\emptyset$.

{\bf Lemma 3} i) $E_x^\mu\cap E_y^\nu\ne\emptyset$ iff
$\Delta_{\mu\cdot\nu}\ge0$; ii) $\mathbb E_x^\mu\subset \mathbb
E_y^\nu$ infers $\mathbb E_y^{\bar\nu}\subset \mathbb
E_x^{\bar\mu}$; iii) $(\exists \mu,\nu)\ E_x^\mu\cap
E_y^\nu\cap\mathbb E\ne\emptyset$ iff either $R\ge 0$, or $(\exists
\mu,\nu,\tau)\ D_{\mu\nu}\ge0$ and $\Delta_\tau\ge0$; iv) $R\ge0$
iff $\vec g\in \mathbb E$; v) $(\forall \mu)\ \Delta_{\mu}<0$ infers
$(\exists\mu,\nu)\ D_{\mu\nu}>0$; vi) Provided $R<0$,
 $(\forall\mu,\nu)\ D_{\mu\nu}<0$
iff $|\alpha|<1$ and $|\beta|<1$.

{\bf Proof }  i) Consider the straight line passing through two
points $Q_{\bar\nu\mu}$ and $Q_{\nu\bar\mu}$ (dashed line in Fig.~2
for the case of $\mu=\nu=+$). If $\Delta_{\mu\cdot\nu}\ge 0$ then
one intersection of $E_x^\mu$ (or $E_y^\nu$) with the straight line
will not lie in the interior of $\mathbb E_y^\nu$ (or $\mathbb
E_x^\mu$ respectively) which means neither $\mathbb E_x^\mu\subset
\mathbb E_y^\nu$ nor $\mathbb E_y^\nu\subset \mathbb E_x^\mu$ and
hence $E_x^\mu\cap E_y^\nu\ne\emptyset$. If $\Delta_{\mu\cdot\nu}<
0$ then, e.g., $A_\mu- B_\nu>|\vec m-\mu\nu\vec n|$ and $\vec z\in
E_y^\nu$ infers $|\vec z-\vec q_{\nu\mu}|+|\vec z-\vec
q_{\bar\nu\mu}|\le 2B_\nu+2|\vec m-\mu\nu\vec n|< 2A_\mu$, i.e.,
$E_x^\mu\cap E_y^\nu=\emptyset$.

ii) $\mathbb E_x^\mu\subset \mathbb E_y^\nu$ is equivalent to
$\Delta_{\mu\cdot\nu}\le 0$, i.e., $\Delta_{\bar\mu\cdot\bar\nu}\le
0$, and $A_\mu\le B_\nu$, i.e., $B_{\bar\nu}\le A_{\bar\mu}$.

iii) Suppose $\vec z\in E_x^\mu\cap E_y^\nu\cap\mathbb E$ for some
$\mu,\nu=\pm$. Since $\vec z\in E_x^\mu\cap E_y^\nu$ we have
$\Delta_{\mu\cdot\nu}\ge0$, $r+|\vec r +2\nu\vec m|=2A_\mu$ and
$r+|\vec r+2\mu\vec n|=2B_\nu$ where $\vec r=\vec z-\vec q_{\nu\mu}$
and $r=|\vec r|$. It follows that $\vec s\times \vec r=\vec
K_{\mu\nu}-r \vec L_{\mu\nu}$  whose square provides a quadratic
equation of $r$:
$(L^2_{\mu\nu}-s^2)r^2-2r\vec K_{\mu\nu}\cdot\vec
L_{\mu\nu}+K^2_{\mu\nu}=0$
where $\vec K_{\mu\nu}=\nu a_\mu\vec n-\mu b_\nu\vec m$,
$L_{\mu\nu}=|\vec L_{\mu\nu}|$ and $K_{\mu\nu}=|\vec K_{\mu\nu}|$.
By noticing $L^2_{\mu\nu}> s^2$ as long as $s>0$ we obtain two
solutions
\begin{eqnarray}\label{r}
r^{(\pm)}_{\mu\nu}
=d_{\mu\nu} +\frac{s^2D_{\mu\nu}\pm \sqrt{s^2a_{\mu}
b_{\nu}\Delta_{\mu\cdot\nu}}}{L^2_{\mu\nu}-s^2}
\end{eqnarray}
and we denote $E_x^\mu\cap E_y^\nu=\{\vec z^{\; (+)}_{\mu\nu},\vec
z^{\; (-)}_{\mu\nu}\}$ with $\vec z^{\; (\pm)}_{\mu\nu}=\vec
q_{\nu\mu}+\vec r^{\; (\pm)}_{\mu\nu}$ and $s^2\vec r^{\;
(\pm)}_{\mu\nu}=(\vec K_{\mu\nu}-r^{(\pm)}_{\mu\nu}\vec
L_{\mu\nu})\times \vec s$.
 The condition
$(\exists\tau)\ \vec z^{\; (\tau)}_{\mu\nu}\in\mathbb E$, i.e., $
2(A_\mu+B_\nu)-r^{(\tau)}_{\mu\nu}+|\vec r^{\;
(\tau)}_{\mu\nu}+2\vec q_{\nu\mu}|\le 4, $ is equivalent to
$(\exists\tau)\ r^{(\tau)}_{\mu\nu}\ge
d_{\mu\nu}-\min\{d_{\bar\mu\bar\nu},0\}$. Due to
\begin{equation}\label{R}
s^2a_\mu
b_\nu\Delta_{\mu\cdot\nu}=s^4D_{\mu\nu}^2+s^2R(L^2_{\mu\nu}-s^2)
\end{equation}
 and Eq.~(\ref{r}), it follows from $(\exists\tau)\ r^{(\tau)}_{\mu\nu}\ge d_{\mu\nu}$ that either
$R\ge0$, or $R<0$ and $D_{\mu\nu}\ge0$. Necessity is thus proved.

If $ \Delta_{\pm}\ge0$ then $(\forall\mu,\nu)\ d_{\mu\nu}\ge0$ since
$2d_{\mu\nu}=\Delta_{\mu\cdot\bar\nu}+a_\mu+b_\nu$. Thus from
$(\exists\mu,\nu)\ D_{\mu\nu}\ge0$ and $R<0$ it follows that
$r_{\mu\nu}^{(\pm)}\ge d_{\mu\nu}$ which infers $\vec z^{\;
(\pm)}_{\mu\nu}\in\mathbb E$ so that $E_x^\mu\cap E_y^\nu \cap
\mathbb E\ne\emptyset$.

If $(\exists\tau)\ \Delta_\tau<0$ and $\Delta_{\bar\tau}\ge0$ then
$(\forall \nu)\ E_x^\nu\cap E_y^{\tau\cdot\nu}=\emptyset$ and
$(\forall \nu)\ E_x^\nu\cap E_y^{\bar\tau\cdot\nu}\ne\emptyset$. It
follows that either $\mathbb E_x^\nu\subset \mathbb
E_y^{\tau\cdot\nu}$ or $\mathbb E_y^{\tau\cdot\nu}\subset \mathbb
E_x^\nu$, i.e., either $\mathbb E_y^{\tau\cdot\bar\nu}\subset
\mathbb E_x^{\bar\nu}$ or $\mathbb E_x^{\bar\nu}\subset \mathbb
E_y^{\tau\cdot\bar\nu}$. As a result either $\mathbb J=\mathbb
E_x^\nu\cap\mathbb E_y^{\tau\cdot\bar\nu}$ or $\mathbb J=\mathbb
E_x^{\bar\nu}\cap\mathbb E_y^{\tau\cdot\nu}$ from which it follows
that $(\exists \nu)$ $E_x^\nu\cap E_y^{\tau\cdot\bar\nu}\subset
\mathbb E$, i.e., $(\exists \tau,\nu)\ E_x^\nu\cap
E_y^{\tau\cdot\bar\nu}\cap\mathbb E\ne\emptyset$.

If $R\ge0$ then we claim that $\Delta_\pm\ge 0$, from which it
follows immediately that $(\forall \mu,\nu)\ E_x^\mu\cap
E_y^\nu\ne\emptyset$ and $\vec z^{\; (+)}_{\mu\nu}\in \mathbb E$.
Firstly, if $a_\pm=0$ (or $b_\pm=0$) then $R\ge0$ infers $s=0$,
which is precluded. Secondly,  if either $(\forall \mu,\nu)\
a_{\mu}b_{\nu}>0$, or $(\exists \mu)\ a_\mu=0$ and $a_{\bar\mu}>0$
and $b_\pm>0$, or $(\exists \nu)\ b_\nu=0$ and $b_{\bar\nu}>0$ and
$a_\pm>0$, then the claim is obviously true due to  identity
Eq.~(\ref{R}).  Thirdly, if $(\exists \mu,\nu)\ a_\mu=b_\nu=0$ and
$a_{\bar\mu}b_{\bar\nu}>0$ then $R=0$,
$D_{\mu\nu}=D_{\mu\bar\nu}=D_{\bar\mu\nu}=0$ with
$D_{\bar\mu\bar\nu}=-4$, and
$\Delta_{\mu\cdot\nu}=\Delta_{\bar\mu\cdot\bar\nu}>0$. As a result
$r_{\mu\nu}^{(\pm)}=d_{\mu\nu}\ge0$ which leads to
$\Delta_{\mu\cdot\bar\nu}=2d_{\mu\nu}\ge0$.

iv) If $R\ge0$ then $(\forall \mu,\nu)\ d_{\mu\nu}\ge0$ so that
$(\forall \mu,\nu)\ |\vec g-\vec q_{\nu\mu}|\le d_{\mu\nu}$, which
infers $\vec g\in\mathbb E$. If $\vec g\in\mathbb E$ then
$(\exists\mu)\ \vec g\in \mathbb E_x^\mu$. As a result $a_\mu-A_\mu
d_{\mu+}=(\vec g-\vec q_{+\mu})\cdot\vec m \le a_\mu-|\vec g-\vec
q_{+\mu}|A_\mu$ which infers either $|\vec g-\vec q_{+\mu}|\le
d_{\mu+}$, i.e., $R\ge 0$, or $A_\mu=0$ which leads to $R=4y^2\ge0$.

v) $\Delta_\pm<0$ infers $R<0$, i.e.,
$(1\pm\gamma)^2<\Delta_{\mp}+|\vec g|^2$, and thus $|\vec
g|>1+|\gamma|$. Let $\eta=\mbox{sgn}[\beta]$ and
$\tau=\mbox{sgn}[\alpha]$ then $|\vec g|\le
A_\eta|\alpha|+B_\tau|\beta|\le D_{\eta\tau}+1+|\gamma|$ which means
$D_{\eta\tau}>0$.

vi) If $(\forall \mu,\nu)\ D_{\mu\nu}<0$ then $|\alpha|<1$ and
$|\beta|<1$ since $\max\{D_{-+},D_{+-}\}+\max\{D_{--},D_{++}\}<0$.
If $|\alpha|<1$ and $|\beta|<1$ then $|\vec g-\vec q_{\nu\mu}|\le
A_\mu(1-\nu\alpha)+B_\nu(1-\mu\beta)$ which, together with
$(\forall\mu,\nu)\ d_{\mu\nu}<|\vec g-\vec q_{\nu\mu}|$ inferred
from $R<0$, leads to $(\forall\mu,\nu)
D_{\mu\nu}=d_{\mu\nu}-A_\mu(1-\nu\alpha)-B_\nu(1-\mu\beta)<0$.

{\bf Proof of Theorem 2 } From Lemmas 1 and 2 and statements i),
iii), and v) of Lemma 3 it follows that two observables are jointly
measurable iff either  $R\ge 0$ or $(\exists\mu,\nu)\
D_{\mu\nu}\ge0$ and Theorem 2 is an immediate result of statements
iv) and vi) of Lemma 3.

{\bf Proof of Theorem 3 } a) $R\ge 0$ is equivalent to $(\forall
\mu,\nu)\ |\vec g-\vec q_{\nu\mu}|\le d_{\mu\nu}$, which means
$(\forall \mu,\nu)\ M_{\mu\nu}(\gamma,\vec g)\ge0$.

b) From $\Delta_\pm\ge0$ it follows that $(\forall \mu,\nu)\
E_x^\mu\cap E_y^\nu\ne\emptyset$ and $ d_{\mu\nu}\ge 0$ and from
$\max\{|\alpha|,|\beta|\}\ge1$ and the choice of $\eta,\tau$ as in
Theorem 3.b it follows that $D_{\eta\tau}\ge0$. As a result $\{\vec
z_{\eta\tau}^{\;(\pm)}\}=E_x^\eta\cap E_y^\tau\subset\mathbb E$ so
that $\vec z_{\eta\tau}^{\;(\pm)}\in\mathbb J$ (Lemma 2). Since
$\mathbb J$ is convex we obtain $\vec z_{\eta\tau}=(\vec
z_{\eta\tau}^{\;(+)}+\vec z_{\eta\tau}^{\;(-)})/2\in \mathbb J$ and
$M_{\mu\nu}(Z(\vec z_{\eta\tau}),\vec z_{\eta\tau})$ is a joint
observable (Lemma 1).

From now on $s$ may be 0. For simplicity we denote by $\Pi_i$
($i=1,2,3,4$) four functions $s^2(\pm\alpha-1)$ and
$s^2(\pm\beta-1)$ and $\Pi=\max_i\{\Pi_i\}$. A set of iff conditions
for joint measurement reads $s^2R\ge 0$ or $\Pi\ge0$. We have

{\bf Lemma 4} $\Pi=0$ infers $s^2R\ge0$.

{\bf Proof} a) If $s>0$ then $\Pi=0$ infers
$\max\{|\alpha|,|\beta|\}=1$, e.g., $|\alpha|=1$ and $|\beta|\le 1$.
Thus $|\vec g-\vec q_{\nu\mu}|=(1-\mu\beta)n\le (1-\mu\beta)B_\nu\le
d_{\mu\nu}$ which is exactly $R\ge0$. Here $\nu=\mbox{sgn}[\alpha]$
and $\mu={\rm sgn}[B_\nu\beta +\nu\gamma-x]$. b) If $s=0$ then
$\Pi=0$ infers $s^2\alpha=s^2\beta=0$ and thus $s^2R=0$.

{\bf Proof of Theorem 1 } We have only to prove that Eq.~(\ref{one})
is equivalent to either $s^2R\ge0$ or $\Pi\ge0$. From the identity
$s^2R=(\gamma^2-f_-)(f_+-\gamma^2)$, where $f_-$ is the l.h.s. of
Eq.~(\ref{one}) and $f_+=f_-+\sqrt{a_+a_-b_+b_-}$, it follows that
$s^2R\ge0$ is equivalent to $f_-\le\gamma^2\le f_+$. Thus we have
only to show that $\gamma^2\ge f_+$ infers $\Pi\ge 0$ and that
$\Pi\ge0$ infers $\gamma^2\ge f_-$. We notice first of all that
$\Pi_i$ are four quadratic (or linear) functions of $c=\vec
m\cdot\vec n$ by regarding $x,y,m,n$ as parameters and  $\Pi$ is
continuous. Case a) $F_x^2+F_y^2\le 1$. In this case $mn\ge |xy|$
and $f_\pm\ge 0$ and $\Pi\le 0$ for $c=xy$ since $|y|\le F_y^2$ and
$F_y^2(n^2-x^2)\le m^2n^2-x^2y^2$. Now that $\Pi\ge 0$ for $c=\pm
mn$, there exist $-mn\le c_-\le xy\le c_+\le mn$ such that $\Pi=0$
for $c=c_\pm$, which infers $xy\pm\sqrt{f_{\mp}}\le c_\pm\le
xy\pm\sqrt{f_\pm}$ (Lemma 4). If $\gamma^2\ge f_+$ then $c\le c_-$
or $c\ge c_+$, which ensures $\Pi\ge0$ since  all the coefficients
of $c^2$ of $\Pi_i$ $(i=1,2,3,4)$ are nonnegative. If $\Pi\ge0$ then
$c\le c_-$ or $c\ge c_+$, which infers $\gamma^2\ge f_-$. Case b)
$F_x^2+F_y^2\ge1$. In this case $\gamma^2\ge f_-$ always and we have
only to show that $\gamma^2\ge f_+$ infers $\Pi\ge0$. If $\Pi=0$ has
no solution then $\Pi>0$ for all $c$ since $\Pi>0$ for $c=\pm mn$.
Let $c_-\le c_+$ be its two solutions and it follows that
$(c_\pm-xy)^2\le f_+$. As a result if $\gamma^2\ge f_+$ then $c\ge
c_+$ or $c\le c_-$, which ensures $\Pi\ge 0$.

{\em Remarks --- } We have derived a single inequality as the
condition for the joint measurement of two simple qubit observables,
based on which an example of MUR is established that generalizes the
existing results.  On finishing this work two references
\cite{b2,srh} provide two seemingly different solutions to the same
problem considered here, whose equivalency can be established in an
analytical or a half-numerical and half-analytical way (see
Appendix) via our results.  The problems of the joint measurability
of more than two observables or observables with more than 2
outcomes are left open.

{\em Acknowledgement --- } This work is supported by the NNSF of
China,
the CAS, the Anhui Provincial Natural Science
Foundation (Grant No. 070412050) and the A*STAR grant
R-144-000-189-305.

\newpage

\paragraph{Comparison with known results--- }Here we shall formulate
those results in \cite{b2,srh} in our notations and and examine the
boundary of admissible $\vec n$ by fixing $y,x,m$. The same boundary
means the equivalency.

{\bf SRH Theorem }\cite{srh} Two observables $\mathcal O(x,\vec m)$
and $\mathcal O(y,\vec n)$ are jointly measurable iff either
\begin{itemize}
\item[(C1)]$\sqrt{1-|y|}\le F_x$;
or
\item[(C2)]$\sqrt{1-|y|}>F_x$ and
$|\gamma|\ge l$; or
\item[(C3)]$\sqrt{1-|y|}>F_x$ and
$|\gamma|<l$ and $\sqrt{a_+h_-}+\sqrt{a_-h_+}\ge 2s$.
\end{itemize}
Here $s=|\vec m\times\vec n|$, and $\gamma=\vec m\cdot\vec n-xy$,
and $a_\pm=(1\mp x)^2-m^2$, and $h_{\pm}=m^2-(\gamma\pm y)^2$, and
$$l=\sqrt{y^2+m^2-|y|(1-x^2+m^2)}.$$

Remarks. The corresponding boundary is plotted in Fig.~3(a) (with
the same parameters as in Fig.~1(a)). If (C1) then $F_x^2+F_y^2\ge
1$ so that Eq.~(\ref{one}) holds always true. If (C2) then, by
noticing that the left-hand-side $f_-$ of Eq.~(\ref{one}) can be
rewritten as
\begin{equation}
f_-=\frac{(a_++2x)(b_++2y)-\sqrt{a_-a_+b_-b_+}}2+m^2+n^2-1,
\end{equation}
we have $f_-\le l$ so that Eq.~(\ref{one}) holds true. If
$1-|y|>F_x^2$ and $|\gamma|< l$ then $|\gamma|< m-|y|$ so that Lemma
4.a applies and Eq.~(\ref{srh}) coincides with Eq.~(\ref{one}). Thus
we have reproduced the boundary in \cite{srh} analytically.

{\bf BS Theorem }\cite{b2} Two observables $\mathcal O(x,\vec m)$
and $\mathcal O(y,\vec n)$ are jointly measurable iff either
\begin{itemize}
\item[(53)] $4\Delta_+ s^2\le a_+b_+(\vec L^2_{--}-s^2)$; or
\item[(54)] $4\Delta_+ s^2\le a_-b_-(\vec L^2_{++}-s^2)$; or
\item[(55)] $4\Delta_+ s^2\le 2(A_+B_+-c)(A_-B_--c)(s^2-\vec L_{++}\cdot\vec
L_{--})\\
-(A_+B_+-c)^2(\vec L^2_{--}-s^2)-(A_-B_--c)^2(\vec L^2_{++}-s^2)$.
\end{itemize}
Here $s=|\vec m\times\vec n|$, and $\Delta_+=(\vec m-\vec
n)^2-(x-y)^2$, $a_\pm=(1\mp x)^2-m^2$, and $b_\pm=(1\mp y)^2-n^2$,
and,
$$\vec L_{\mu\mu}=\mu(1-\mu x)\vec n-\mu(1-\mu y)\vec m,$$
and $A_\pm=1\mp x$ and $B_\pm=1\mp y$ and $c=\vec m\cdot\vec n$.

Remarks.  Despite the facts that we have identified Eq.~(55) with
$R\ge0$ (Lemma 4.b) and that the boundaries $R=0$ and $|y|+n=1$ and
$$4\Delta_+ s^2=\max_{\mu=\pm}\{ a_\mu b_\mu(\vec
L^2_{\bar\mu\bar\mu}-s^2)\}$$ intersect at exactly where
$\max\{|\alpha|,|\beta|\}=1$ and that numerical evidences indicate
that BS conditions also give rise to the same boundary, we fail to
work out an analytical proof for the equivalency so far. The
corresponding boundary is plotted in Fig.~3(b).  The red-contoured
region comes from $R\ge 0$ while the blue-contoured region comes
from the conditions Eqs.(53,54).

{\bf Lemma 5} a) Either $R\ge0$ or $\{|\beta|\ge1$ and $h_\pm\ge0$\}
iff
\begin{equation}\sqrt{a_+h_-}+\sqrt{a_-h_+}\ge 2s.\label{srh}\end{equation}
b) Condition Eq.~(55) is equivalent to $R\ge0$.

{\bf Proof } a) If $R\ge0$ then $\vec g\in \mathbb E_x^+\cap\mathbb
E_x^-$ so that $h_\pm\ge0$ and $|(1\pm\beta)s|\le\sqrt{a_\mp h_\pm}$
which infers Eq.~(\ref{srh}). If $|\beta|\ge1$ and $h_\pm\ge0$,
since $4\beta s^2=h_+a_--h_-a_+$, then $ 4s^2\le h_+a_-+h_-a_+$ and
Eq.~(\ref{srh}) follows. On the other hand if
 Eq.~(\ref{srh}) holds true then
$h_\pm\ge0$ and $(\exists\mu)\ (1-\mu\beta)|s|\le\sqrt{a_\mu
h_{\bar\mu}}$ which infers either $|\beta|\ge1$ or $\vec g\in\mathbb
E_x^\mu$, i.e., $R\ge0$.

b) It follows from the identities $A_+B_++A_-B_--2c=2(1-\gamma)$ and
$(A_+B_+-c)\vec L_{--}+(A_-B_--c)\vec L_{++}=2(y+\gamma x)\vec
n-2(x+\gamma y)\vec m$ whose length squared is equal to $4s^2|\vec
g|^2$ and $R=(1-\gamma)^2-|\vec g|^2-\Delta_+$.

\begin{figure}
\includegraphics[width=40mm]{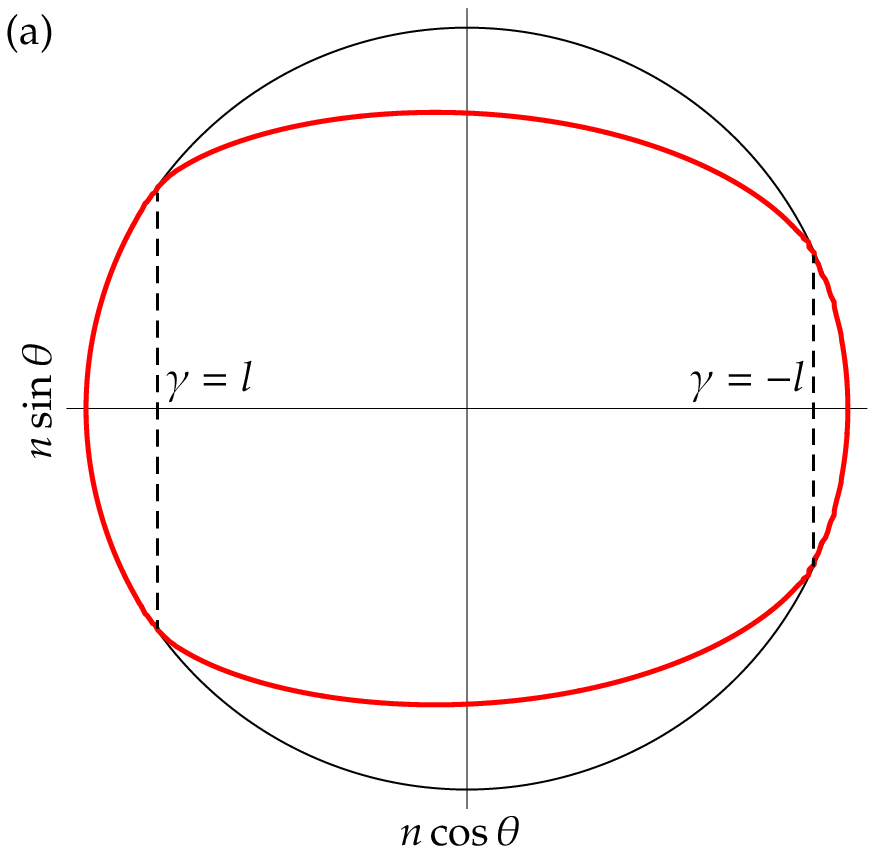}\hfill\includegraphics[width=40mm]{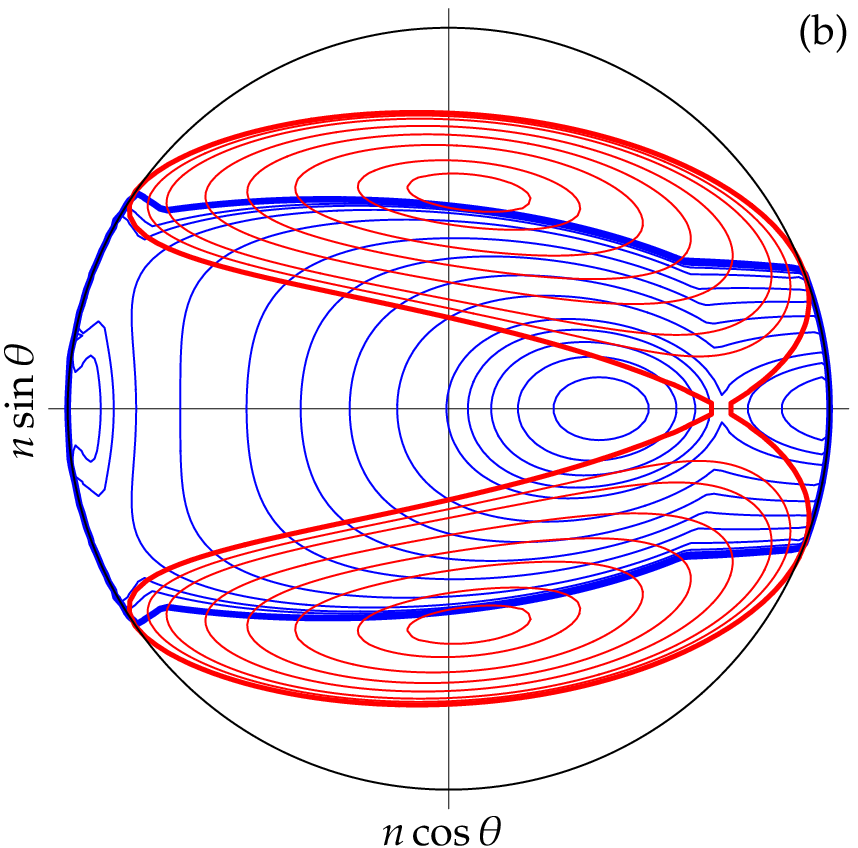}
\caption{The boundary of admissible $\vec n$ arising from (a) SRH
conditions; (b) BS conditions with fixed $m=0.8,x=-0.1$, and
$y=0.3$}
\end{figure}

\paragraph{Conclusion--- } We have proved the equivalency between SRH
conditions \cite{srh} and ours analytically and the equivalency
between BS conditions \cite{b2} and ours (so that SRH conditions)
half-analytically and half-numerically.
\newpage
\end{document}